\documentclass[conference]{IEEEtran}
\usepackage{amsmath,amssymb,multirow}

\usepackage{graphicx,tikz,scalefnt,verbatim,balance,url}
\usepackage[ruled,linesnumbered,nofillcomment,noline]{algorithm2e}
\SetCommentSty{textrm}
\usepackage[font=small]{subfig}

\title{Trimming the Multipath\\ for Efficient Dynamic Routing}

\author{
\IEEEauthorblockN{Adrian S.-W. Tam \hspace{2em}
Kang Xi \hspace{2em}
H. Jonathan Chao}\\
\IEEEauthorblockA{Department of Electrical and Computer Engineering\\
Polytechnic Institute of
New York University\\
Email: adriantam@nyu.edu, kxi@poly.edu, chao@poly.edu}
}

\begin{document}

\maketitle

\begin{abstract}
Multipath routing is a trivial way to exploit the path diversity to leverage
the network throughput. Technologies such as OSPF ECMP use all the available
paths in the network to forward traffic, however, we argue that is not
necessary to do so to load balance the network. In this paper, we consider
multipath routing with only a limited number of end-to-end paths for each
source and destination, and found that this can still load balance the traffic.
We devised an algorithm to select a few paths for each source-destination pair
so that when all traffic are forwarded over these paths, we can achieve a
balanced load in the sense that the maximum link utilization is comparable to
that of ECMP forwarding. When the constraint of only shortest paths (i.e. equal
paths) are relaxed, we can even outperform ECMP in certain cases. As a result,
we can use a few end-to-end tunnels between each source and destination nodes
to achieve the load balancing of traffic.
\end{abstract}

\section{Introduction}\label{sec:intro}
The networks nowadays, in particular the data center networks, have a high
degree of connectivity and redundancy. The traditional shortest path routing
does not make use of them for a better performance. Such routing paradigm
forward packets from its source to the destination on only one path. This may
lead to imbalanced load and create `hot-spot' links. Introducing multipath
routing, such as OSPF \cite{rfc2328} ECMP, is an easy way to leverage the
overall throughput of the network as well as balancing the load on different
links.

There are two categories of multipath routing mechanisms in the current
literature: hop-by-hop routing or end-to-end tunnels. The former, OSPF ECMP as
an example, is scalable as each intermediate router selects the next hop of a
route independently, and the routing tables on the routers are built
distributed. The latter mechanism is to build parallel end-to-end tunnels
between a pair of end-points, such as using MPLS label switched paths
\cite{rfc3031}. Then the traffic between the two end-points are forwarded over
these established tunnels. This mechanism allows the forwarding paths to be
known explicitly. It also allows the network administrators to have more
control on the traffic over the network. For example, we can build a particular
tunnel to avoid traversing certain part of the network; also we can adjust the
proportion of the traffic sending over different tunnels to precisely control
the load on different links.

The controllability provided by the multipath routing using end-to-end tunnels
are attractive to the network administrators as it is easier to load-balance
the networks. However, there are hard limitations on the number of tunnels you
can built. For example, SPAIN \cite{myam10} uses VLANs to separate the parallel
forwarding paths, then the number of VLANs that it can use is subject to the
size of VLAN ID space; MATE \cite{ejlw01} builds upon MPLS, therefore it is
subject to the limitation of MPLS label space. Moreover, on the same
source-destination pair, the more the number of parallel paths, the higher the
overhead in operation, such as the time and bandwidth spent on probing realtime
path characteristics, and the computation resource on finding the optimal
traffic splitting ratio among the paths.

In this paper, we consider a data center network built with modern switches
such as OpenFlow switches \cite{mabpprst08}, similar to the case in
\cite{baaz10}. In order to have a higher network throughput and prevent
congestion, we forward traffic on multipath routes. The multipath routes are
prebuilt end-to-end connections for the flexibility in control. Because of the
delay sensitivity in data center applications, shortest paths between the
source and destination nodes are preferred. And because of the overhead in
maintaining these connections, we would like to keep the number of parallel
paths between a pair of nodes to be as small as possible, while allowing the
network to be load-balanced and flows routed dynamically. In this setting, we
devised an algorithm to prebuild the multipath for all possible connections
with the objective to minimize the network congestion, namely, to minimize the
maximum link load on the network. We do not exploit all the multipath
possibility provided by the network, but use a subset of the multipath to
forward traffic. The contributions of this paper are (1) providing a method to
strategically select a few paths for multipath forwarding, and (2) showing that
even the number of paths used is few, we can achieve a performance similar or
even better than ECMP, which forward traffic over \emph{all} equally shortest
paths.

\section{Related Work}\label{sec:related}

Multipath routing have been discussed since the early years of networking. A
large body of literature is available addressing different issues of multipath
routing. For example, \cite{ns99,vg00,vg01,mpc08} describe the way to deduce
loop-free multipath routes in a hop-by-hop basis. Some other, such as
\cite{ft00,bo07}, transform the multipath routing problem into constrained
optimization to find the optimal way to forward traffic for highest throughput
and lowest congestion. QoS routing is addressed in
\cite{wc96,ms97,rfc2386,cn98,jng01}, which is to perform multipath routing to
satisfy QoS guarantees.

Multipath has the potential to allow dynamic load balancing without route
change. There are proposals on splitting traffic to different paths adaptive to
the current network conditions for a better load balance, such as the OSPF
Optimized Multipath \cite{ospfomp} and MPLS Optimized Multipath \cite{mplsomp}. 
MATE \cite{ejlw01} proposed an analytical model and algorithms to adjust the
traffic ratio on the multipaths in order to reduce the network congestion.
Similarly, in \cite{nzd04}, the authors propose to use multipath routing
adaptively by adjusting the traffic split ratio among different path
according to regularly updated utilization information. The context of
\cite{nzd04} is in QoS routing, i.e. bandwidth on the path is explicitly
reserved for a flow. However, it also recognizes the need to reduce the number
of paths being used due to overhead concern. An algorithm is proposed in
\cite{nzd04} to find the \emph{widest disjoint paths} for a source-destination
pair, so that it provides the highest probability of satisfying the QoS
guarantee upon a flow arrival. It share the same objective as our work but we
are different in certain sense: Firstly, the multipath find in \cite{nzd04}
depends on the current utilization information. Such statistics is volatile and
the selection of the multipath is updated time to time. Our work, however, is
based on an estimate of the traffic matrix to find the multipath. The multipath
is stable amid the fluctuating utilization on the network. The insensitivity of
our performance to the traffic matrix is evaluated in section \ref{sub:varied}.
Secondly, the multipath is selected individually on each sender node in
\cite{nzd04}. This approach does not allow the coordinated multipath selection,
therefore it may accidentally lead the concurrent flows to contention at
certain links. Our approach is to do the multipath selection from a centralized
view, so that by strategically selecting different paths for the traffic, the
load-balancing objective is achieved.

\section{Problem Statement}\label{sec:problem}

The easiest way to run multipath routing in commodity switches and routers is
to turn on the equal-cost multipath forwarding (ECMP). ECMP performs per-hop
decision to evenly distribute traffic to all the next hops that are equally
close to the destination. Because of its ubiquitous availability, ECMP
forwarding is also proposed in data center networks to obtain a higher
throughput. See VL2 \cite{gjkklmps09} for example. This is a multipath routing
using hop-by-hop decisions. Therefore, the forwarding paths of a flow according
to ECMP may not be mutually disjoint, and the traffic may not be well balanced.

\begin{figure}
\begin{center}
\begin{tikzpicture}[scale=1.33]
\begin{scope}[line width=1pt]
\node[circle,draw] (S) at (0.0,1.0) {S};
\node[circle,draw] (T) at (4.0,1.0) {T};
\node[circle,draw] (A) at (1.0,2.0) {A};
\node[circle,draw] (B) at (1.0,0.0) {B};
\node[circle,draw] (C) at (3.0,2.0) {C};
\node[circle,draw] (D) at (3.0,0.0) {D};
\draw (S) --node[anchor=north,rotate=-45]{0.5} (B);
\draw (A) -- (B);
\draw (B) --node[anchor=north]{0.25} (D);
\draw (C) -- (D);
\draw[->] (S) --node[anchor=south,rotate=45]{0.5} (A);
\draw[->] (A) --node[anchor=south]{0.5} (C);
\draw[->] (B) --node[anchor=south,rotate=45]{0.25} (C);
\draw[->] (C) --node[anchor=south,rotate=-45]{0.75} (T);
\draw[->] (D) --node[anchor=north,rotate=45]{0.25} (T);
\end{scope}
\end{tikzpicture}
\end{center}
\caption{ECMP forwarding one unit of traffic from S to T leads to imbalanced traffic load.}\label{fig:badecmp}
\end{figure}
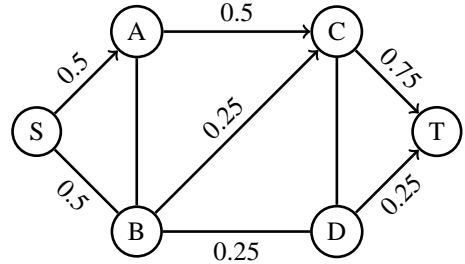

\figurename~\ref{fig:badecmp} shows an example that the parallel
paths according to ECMP are not disjoint and the traffic is imbalanced. We
assume each link in \figurename~\ref{fig:badecmp} has equal weight in the
shortest path computation. The arrows between nodes denotes the shortest path
tree toward T. When one unit of traffic is sent from S to T, the ECMP routing
logic will split the traffic to each link as shown in the figure. Here we can
see, the link CT is carrying three times more traffic than link DT, because at
each node ECMP splits traffic equally among all the valid next hops. To solve
this problem, optimized multipath \cite{ospfomp} and adaptive multipath
\cite{gzrr03} are proposed. They make use of the packet loss information to
split the traffic among different paths. In essence, the traffic load on a
path is adaptive to the level of congestion. However, we can see that, the same
problem can be solved easily by removing the path S-B-C-T, i.e. not to use all
the available multipaths but a subset of them.

How can we use a subset of available paths in multipath routing? An easy way is
to build the multipath using tunnels, such as the MPLS label switched paths. As
the OpenFlow architecture \cite{mabpprst08} getting popular, we can also make
use of its flexibility to pin-point a path in OpenFlow switches or manipulating
their forwarding table to achieve the same goal. Therefore, the remaining
problem is that, how should we select the paths for multipath routing, so that
we can have a better load-balancing or equivalently, minimize the congestion.

In this paper, we attempt to solve the following problem: Given a topology, and
a set of flows, find the routes to deliver these flows with the constraint that
no any flow, defined as a traffic demand between a unique source-destination
pair, is routed over more than $k$ paths. For simplicity, we assume the traffic
of a flow is distributed evenly to all the viable paths. In other words, the
traffic distribution is \emph{not} adaptive to network congestions, so that we
avoid the issues of route change and their consequence of packet reordering.
The objective is to route efficiently, namely, to minimize the maximum link
utilization on the network.

This is similar to the multi-commodity flow problem \cite{pm04} but with
additional constraints on splitting traffic evenly to all viable paths.
Afterwards, we will remove the requirement of providing $k$ as a parameter, and
to find the optimal number of paths given the traffic matrix.

\section{The Algorithm}\label{sec:algo}

\subsection{Heuristic algorithm on path selection}
As the complexity in solving the multi-commodity flow problem for the optimal
solution is high, a heuristic algorithm is devised. The heuristic algorithm is
to find the multipaths individually between a pair of nodes. Since our goal is
to balance the network, i.e. minimize the maximum link utilization for the
given traffic matrix, we introduce a cost function to measure how good are the
paths. The outline of the algorithm is depicted in Algorithm \ref{alg:1}.

\begin{algorithm}
\caption{Algorithm to place routes on a network}\label{alg:1}
\KwData{Network topology $G=(V,E)$}
\KwData{Source-destination flows $F \subset V^2$}
\KwData{Estimated traffic matrix $\alpha_{st},\;\forall (s,t)\in F$}
\KwData{Maximum number of paths for a flow $k$}
Initialize link load account $\lambda_e :=0,\quad\forall e\in E$\;
\ForEach{$(s,t)\in F$ in random order}{
  \tcc{Find the candidate paths from $s$ to $t$}
  $M$ := all valid paths joining $s$ to $t$\;\label{alg:1:selectpath}
  \tcc{Select $k$ paths}
  \For{$i:=1$ to $k$}{
    $p$ := the path among $M$ with the lowest path cost\;\label{alg:1:tiebreak}
    $\lambda_e$ increase by $\frac{1}{k}\alpha_{st}$ for all $e\in p$
  }
}
\end{algorithm}

The heuristic algorithm is a greedy algorithm that selects one path at a time,
assuming that we can minimize the overall maximum link utilization by
minimizing the path cost in every step. The algorithm takes a traffic matrix as
the input to provide estimate of the relative size of traffic between different
pairs of nodes. Whether the traffic matrix can be fulfilled in the network is
not a concern, although it may have influence on the cost function, for example, a
cost function that models the links as queues will find an unfulfillable
traffic matrix to yield an infinite cost. The idea of the algorithm is as
follows: We first initialize link load values $\lambda_e$ to zero. They serve
as the scoreboard for the amount of traffic assigned to the links. Then we
examine each pair of nodes in a random order, to select the $k$ best paths on
the network that connect them. A cost function returns a cost for every path
based on the current load on each link. The $k$ best paths are selected so that
they correspond to the lowest cost according to the current $\lambda_e$. Once
the paths are selected, all the corresponding $\lambda_e$ are increased by
$\alpha_{st}/k$, where $\alpha_{st}$ is the amount of traffic from $s$ to $t$,
to reflect that these links are carrying additional traffic. For simplicity, we
assume an even distribution of load to these $k$ paths. This can serve as the
lower bound on the performance if adaptive multipath routing is being used. At
the end of the algorithm, $\lambda_e$ reflects the resulting utilization as the
multipaths are selected accordingly.

The candidate paths may subject to certain constraints. Obviously, it is
reasonable to require these paths to be loop-free. Moreover, we may also limit
the paths' length to avoid a significant disparity in propagation delay between
different paths. These limitations are fulfilled in line
\ref{alg:1:selectpath}. Obviously, if there are no more than $k$ paths
available, all of them would be chosen to route traffic between this pair of
nodes. More detail on the candidate path selection process is presented in next
subsection.

The cost function is the core of this heuristic algorithm. It quantizes the
preference of a path over another. Because our objective is to minimize the
maximum link utilization, we use the following cost function in our paper:
\begin{align}
\chi(p) = \max_{e\in p} \frac{\lambda_e + \alpha_{st}/k}{\mu_e}.
\label{maxfunction}
\end{align}
This function takes a path $p$ as an argument, and find the maximum link
utilization among all the links on this path. The fraction represents the
resulting utilization if this path is chosen to carry traffic from $s$ to $t$,
namely, the total resulting traffic, $\lambda_e+\alpha_{st}/k$ divided by the
link capacity $\mu_e$.

We also tried different cost functions, such as sum of the utilization instead
of maximum, so that it can prefer an overall shorter and less congested path,
or evaluating the cost by a monotonically increasing convex function. However,
according to our evaluation done in section \ref{sec:eval}, the cost function
\eqref{maxfunction} gives the best performance in terms of minimizing the
maximum link utilization.

In line \ref{alg:1:tiebreak}, we use the path length as the tie-breaker if
there are several paths of the equal lowest cost. When these paths are of equal
length, we randomly pick one among them.

\subsection{Path finding algorithm}

In line \ref{alg:1:selectpath} of Algorithm \ref{alg:1}, we enumerate the valid
paths that joins $s$ to $t$. A number of algorithms are available to find not
only a single shortest path but multiple paths between two nodes, for example,
\cite{y71,m78}. In our study, we use a modified Eppstein's algorithm \cite{e98}
for simplicity reasons. Its pseudocode is listed as Algorithm \ref{alg:2}. We
describe the Eppstein's algorithm as follows.

\begin{algorithm}
\caption{Path finding algorithm}\label{alg:2}
\KwData{Network topology $G=(V,E)$}
\KwData{Length of links $w(e)\quad\forall e\in E$}
Run Bellman-Ford algorithm to find the shortest path from any node to $t$,
let $d(v)$ denotes the shortest path distance found from $v$ to $t$\;
$T$ := the tree composed of the the shortest paths to $t$\;
$S$ := the set of links not in $T$, a.k.a. sidetrack links\;
\ForEach{$e\in S$}{
  \tcc{Assume $e$ is a directed edge joining $u$ to $v$}
  $c(e)$ := $d(v) - d(u) + w(e)$
}
Start with an empty priority queue $Q$\;
$Q$.push($\emptyset$ with priority 0)\;
\While{$Q$ not empty}{
  $\sigma$ := $Q$.pop()\;
  Output the shortest path from $s$ to $t$ according to $T$ and $\sigma$\;
  \ForEach{$e\in S-\sigma$}{
    $\sigma'$ := $\sigma \cup \{e\}$\;
    \If{$\sigma'$ represents a valid path}{
      $Q$.push($\sigma'$ with priority $\sum_{e\in\sigma'} c(e)$)\;
    }
  }
}
\end{algorithm}

The core of the Eppstein's algorithm is the shortest path tree $T$ terminated
at $t$ as given by the Bellman-Ford algorithm. The tree is a directed graph
such that each arc is pointed to $t$. We can find the shortest path from any
node $v$ to $t$ by traversing along the arc in this tree. Any loop-free path
other than the shortest path according to this tree $T$ can be represented by a
set of \emph{sidetrack} edges. The sidetrack edges are directed edges. They are
the edges that are traversed in a path that is not included in $T$. When a set of
sidetrack edges are given, the path from $s$ to $t$ is constructed as follows:
We start from node $s$ on the network, and traverse the edges until we reach
$t$. At every node, the edge to traverse is determined uniquely by first
looking for any edge that is included in the set of sidetrack edges. If none
is found, then we follow the edge on the shortest path tree $T$. Therefore, a
well-formed set of sidetrack edges shall not contain more than one edge
emerging from the same node. Moreover, the path constructed in this way shall
exhaust all the provided sidetrack edges to make this set of sidetrack edges
well-formed. Thus, for any loop-free path, there is a corresponding set of
sidetrack edges, but not vice versa.

We use \figurename~\ref{fig:badecmp} to illustrate the idea of the Eppstein's
algorithm. An empty set of sidetrack edges means we traverse along the shortest
path tree. Therefore, for S to T, the path would be S-A-C-T. If the set of
sidetrack edges is A-B and C-D, the path would be S-A-B-C-D-T. But if the set
of sidetrack edges is B-A only, we cannot find a corresponding path since we
can never reach node B from S along the shortest path tree.

The latter half of the path finding algorithm is crucial to enumerate the paths
in ascending order of path length. The quantity $c(e)$ represents the penalty
of using a sidetrack edge $e$ in terms of the length of resulted path, $c(e)\ge
0$ guaranteed. Therefore, for a well-formed set of sidetrack edges,
$\sigma(p)$, the length of the corresponding path $p$ is \[
\ell(p) = d(s) + \sum_{e\in\sigma(p)} c(e) \]
where $d(s)$ is the shortest path distance from $s$ to $t$ and the summation is
the total additional length imposed by traversing the sidetracked edges.
Therefore, for any set $\sigma(p')$ that is a superset of $\sigma(p)$, the
corresponding path of the former must be no shorter than that of the latter.
Making use of this property, the while-loop in the above algorithm use a
priority queue to enumerate the paths in ascending order of lengths. And
because of this, we can terminate the path-finding algorithm prematurely if we
impose a constraint on the length of paths to be found. For example, we can
limit to our search to only shortest paths (i.e. paths of the length equal to
$d(s)$) by skipping the set of sidetrack edges $\sigma(p)$ whenever
$\sum_{e\in\sigma(p)} c(e) > 0$ is found. In our evaluations, we set a
threshold $\theta\ge 0$ to limit the path selection to
\begin{align*}
\frac{1}{d(s)}\sum_{e\in\sigma(p)}c(e) \le \theta,
\end{align*}
i.e. no path that is longer than the shortest path by a fraction $\theta$ will
be selected.

\subsection{Optimal number of paths}\label{sub:optimalk}

According to our evaluation, we find that having more path diversity does not
necessarily decrease the maximum link utilization on the network. Indeed, the
increase in $k$ (the number of paths in a multipath set between a pair of
nodes) exhibits the law of diminishing return. Therefore, it is interesting to
ask, for a particular topology, at least how many paths we needed to enjoy the
benefits of multipath routing.

\begin{algorithm}
\caption{Algorithm to place routes on a network}\label{alg:3}
\KwData{Network topology $G=(V,E)$}
\KwData{Source-destination flows $F \subset V^2$}
\KwData{Estimated traffic matrix $\alpha_{st},\;\forall (s,t)\in F$}
Initialize link load account $\lambda_e :=0,\quad\forall e\in E$\;
Initialize multipath $\pi_{st}:=\emptyset\quad\forall (s,t)\in F$\;
\For{$i:=1$ to $k$}{\label{alg:3:limitk}
  \ForEach{$(s,t)\in F$ in random order}{
    \tcc{Find the candidate paths from $s$ to $t$}
    $M$ := all valid paths joining $s$ to $t$ not in $\pi_{st}$\;
    \tcc{Select the $i$-th path}
    $p$ := the path among $M$ with the lowest path cost\;
    \If{$(\pi_{st}\cup\{p\})$ is better that $\pi_{st}$}{\label{alg:3:optimal}
      \ForEach{path $q \in \pi_{st}$}{
        $\lambda_e := \lambda_e - \frac{1}{|\pi_{st}|}\alpha_{st} + \frac{1}{|\pi_{st}|+1}\alpha_{st}\quad\forall e\in q$\;\label{alg:3:newlambda}
      }
      $\lambda_e := \lambda_e + \frac{1}{|\pi_{st}|+1}\alpha_{st}\quad\forall e\in p$\;
      $\pi_{st} := \pi_{st} \cup \{p\}$\;
    }
  }
}
\end{algorithm}

We slightly modified Algorithm \ref{alg:1} to detect the optimal number of
paths to be used automatically, as outlined in Algorithm \ref{alg:3}. The idea
of finding the optimal number of paths is to compare if it is beneficial to
have one more paths added to the set of multipaths. Since the offered load
$\alpha_{st}$ between a pair $(s,t)$ is given, if it is distributed to $n$
paths, then each path will carry $\alpha_{st}/n$. Obviously, each path will
carry less traffic if $n$ is larger. In other words, the existing links for
$(s,t)$ will benefit from having more path diversity. However, if the traffic
is extended to a new path which involves a heavily loaded link, additional
traffic will make the link even more heavily loaded. In line
\ref{alg:3:optimal} of Algorithm \ref{alg:3}, we check if extending the
existing set $\pi_{st}$ of paths for $(s,t)$ to a new path $p$ is worthwhile by
verifying if the additional path can lower the maximum link load among he
links in concern, i.e.
\begin{align*}
\max \{\lambda_e': e\in q, q\in(\pi_{st}\cup\{p\})\} \le \max \{\lambda_e:e\in q, q\in\pi_{st}\},
\end{align*}
where the new link load $\lambda_e'$ is computed according to the equation in
line \ref{alg:3:newlambda}. If this condition holds, we include $p$ into
$\pi_{st}$ and update the values of $\lambda_e$ correspondingly.

We placed a limit $k$ as the maximum number of paths for a pair of nodes
$(s,t)$, as in line \ref{alg:3:limitk}. In practice, we found that the number
of parallel paths is mostly limited by the condition in line
\ref{alg:3:optimal}.

\subsection{Limitations of the Algorithm}

The algorithms to find the $k$ paths for all the source-destination pairs is
presented in the previous subsections. The algorithms do not try to find the global
optimal but instead, they are heuristic algorithms to approximate the solution.

Because the heuristic algorithms search the paths in a greedy manner, there are
three limitations:

\begin{itemize}
\item As the paths for each source-destination pair is computed
	individually, there is no coordination to strategically use different
	paths for different source-destination pairs to avoid contention. The
	algorithms use only the cost function as the means to avoid such
	contention, which is not guaranteed to success.
\item The way to find $k$ paths for a particular source-destination pair is to
	find the paths one by one according to the updated path cost. In this
	way, similarly, we cannot coordinate the selection between these $k$
	paths.
\item Particular to Algorithm \ref{alg:1}, it finds $k$ paths for a
	source-destination pair whenever it is possible. However, for the
	objective of minimizing the maximum link load, the best solution may
	not be utilizing all the $k$ paths.
\end{itemize}

To partially deal with the limitation of the greedy algorithm and improve the
output, we can fine-tune the path selections after the main algorithm as
follows: Firstly, find the links of the maximum load. This is the set of hot
links in the current configuration. Then, for each path that traverses any hot
links, try to look for a substitute path such that, if replaced, we can lower
the load on one existing hot link but not creating a new hot link. This
algorithm is outlined in Algorithm \ref{alg:4}.

\begin{algorithm}
\caption{Fine-tuning path selection}\label{alg:4}
\KwData{Path set $P$}
\Repeat{no more change in the path selection is possible}{
  \tcc{Find all paths that traverses a hot link}
  $\hat{E} := \arg\max_e \lambda_e$ \;
  $\hat{P} := \{p : p\in P \textrm{ such that }e\in p\ \exists e\in\hat{E}\}$\;
  \ForEach{$p\in\hat{P}$}{
    \tcc{Find a candidate substitution path}
    $(s,t)$ := Source-destination pair of $p$\;
    $\Pi$ := all valid paths joining $s$ to $t$ that (a) not in $P$, and (b) not traversing $\hat{E}$, and (c) can alleviate a hot link without creating a new one\;
    \If{$\Pi\neq\emptyset$}{
      Replace $p$ with any path in $\Pi$\;
      Update $P$
    }
  }
}
\end{algorithm}

Moreover, the algorithm is based on the assumptions that we can use any
arbitrary loop-free paths to forward traffic and the traffic demand is known a
priori. The former assumption can be readily achieved by setting up end-to-end
paths, such as MPLS LSP. If only hop-by-hop forwarding is allowed, the
selection of paths (Algorithm \ref{alg:2}) would be more restricted. The latter
assumption is also a reasonable one as we can merely estimate the traffic
matrix by statistics of historical usage data.

\section{Evaluations}\label{sec:eval}

Evaluations are done on the heuristic algorithm in section \ref{sec:algo}, to
compare its performance against the standard ECMP. The first evaluation is to
compare the traffic load on each links as a result of even distribution
according to ECMP, against using the paths found by the algorithm (section
\ref{sub:subset}).  Subsequently, we investigate the effect of adjusting
different parameters in the algorithm (sections \ref{sub:nonshort} and
\ref{sub:k}) and its sensitivity to traffic matrix (section \ref{sub:varied}).
Finally, we use flow-based simulation to verify our statements on the
performance of the algorithm (section \ref{sub:dyn}).

\subsection{ECMP vs Subset of shortest paths}\label{sub:subset}

We first assume the traffic demand provided is fluid demand, such that, we
can split the traffic evenly across a number of paths. The traffic distribution
as depicted in \figurename~\ref{fig:badecmp} is the result of fluid traffic
splitting according to ECMP. We compare the load distribution according to ECMP
against the case of $k=4$ paths being found by our algorithm with $\theta=0$,
i.e. the paths found are restricted to shortest paths.

\begin{figure*}
\subfloat[]{\includegraphics[width=0.5\columnwidth]{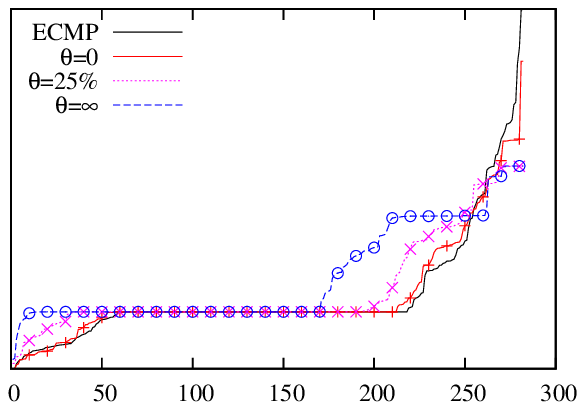}\label{sub:att}}
\subfloat[]{\includegraphics[width=0.5\columnwidth]{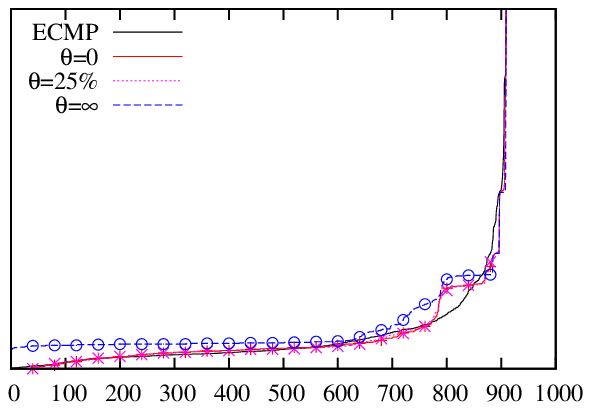}\label{sub:l3}}
\subfloat[]{\includegraphics[width=0.5\columnwidth]{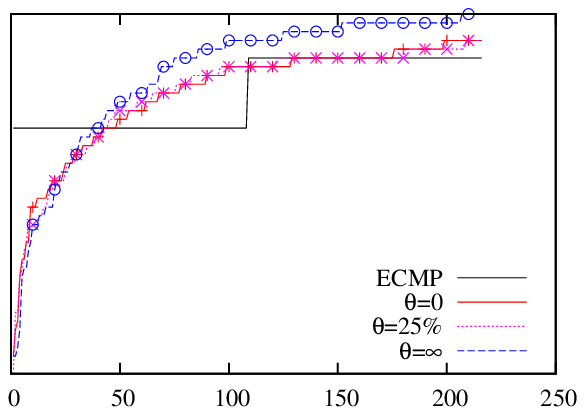}\label{sub:ft3}}
\subfloat[]{\includegraphics[width=0.5\columnwidth]{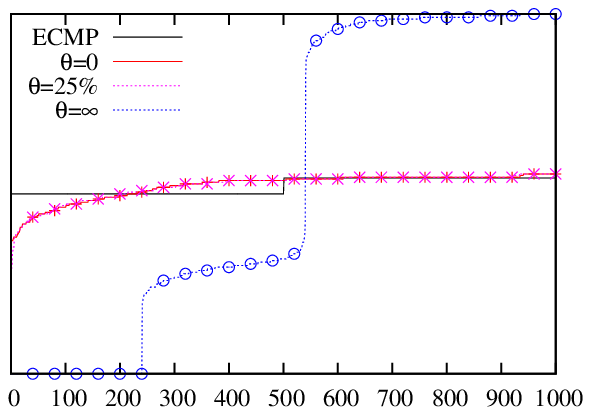}\label{sub:ft5}}
\caption{Relative load of links in \subref{sub:att} irregular topology 1,
\subref{sub:l3} irregular topology 2, \subref{sub:ft3} fat tree
XGFT(2;3,6;3,3), and \subref{sub:ft5} fat tree XGFT(2;5,10;5,5)}\label{fig:const}
\end{figure*}

\figurename~\ref{fig:const} shows the relative link load in different
topologies. It plots the load of links in a topology in ascending order. The
horizontal axes numbers the links; the magnitude of the vertical axes are
unimportant, hence omitted. The plots demonstrate how balanced is the load
when different flow placement strategies are used. Here we assume a uniform
traffic demand, i.e. each distinct pair of hosts in the network sends one unit
of traffic to each other. In irregular topologies (\figurename~\ref{sub:att}
and \ref{sub:l3}), any node can send traffic, but in fat tree topologies
(\figurename~\ref{sub:ft3} and \ref{sub:ft5}), the hosts that can send or
receive traffic are attached to the edge switches.

Consider ECMP in \figurename~\ref{fig:const}, we see that regular topologies
such as fat tree can benefit greatly from ECMP, as their links are highly
balanced (their curves are narrow and flat). Irregular topologies, on the
contrary, has a significant disparity between the heavily loaded and lightly
loaded links. Those heavily loaded links are `hot links' of the network that
are more popular, thus carried more load than others.

When we use our heuristic algorithm to find the multiple shortest paths to
replace ECMP, we found its performance is close to that of ECMP in terms of the
maximum link load in the topology. These two curves are even almost coincide in
irregular topologies and closely approximate to each other in fat trees. The
difference between them is due to the heuristic nature of the algorithm,
namely, it does not coordinate the path selection between different pairs. Even
the two curves are not identical, they are close at the right end. This means
using only \emph{a few} shortest paths to forward traffic is similar to that of
using \emph{all} shortest paths in ECMP, but without sacrificing the balance of
load.

\begin{figure*}
\subfloat[]{\includegraphics[width=0.5\columnwidth]{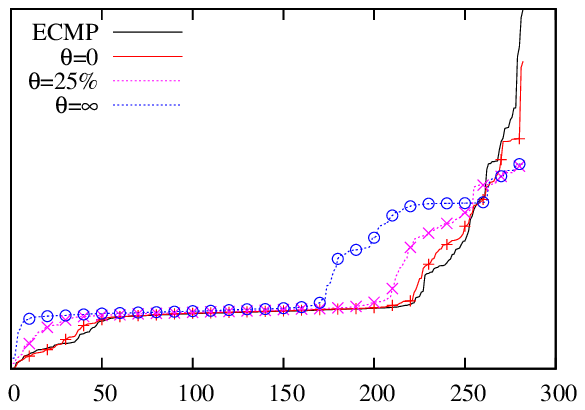}}
\subfloat[]{\includegraphics[width=0.5\columnwidth]{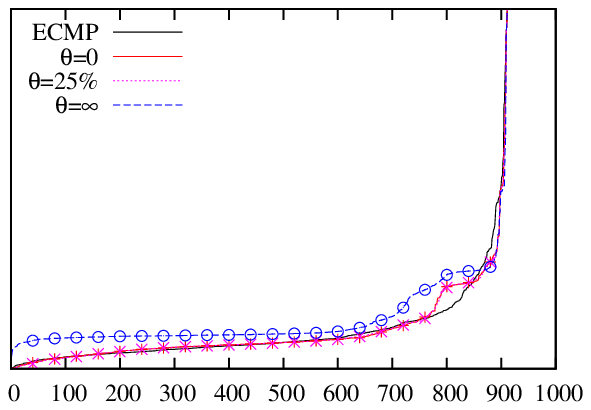}}
\subfloat[]{\includegraphics[width=0.5\columnwidth]{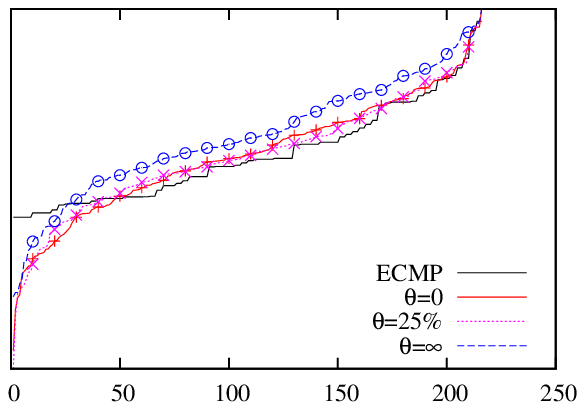}}
\subfloat[]{\includegraphics[width=0.5\columnwidth]{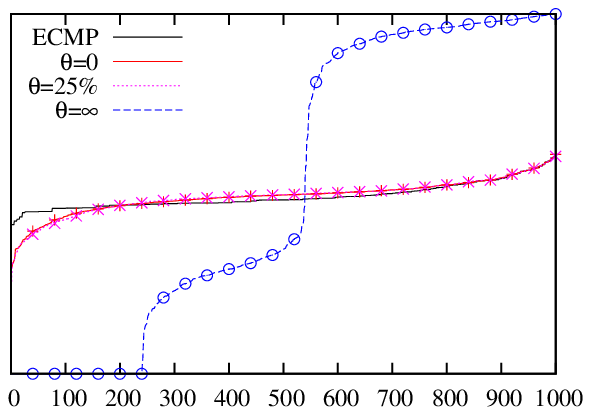}}
\caption{Relative load of links in \subref{sub:att} irregular topology 1,
\subref{sub:l3} irregular topology 2, \subref{sub:ft3} fat tree
XGFT(2;3,6;3,3), and \subref{sub:ft5} fat tree XGFT(2;5,10;5,5). Random traffic demand.}\label{fig:rand}
\end{figure*}

This claim is valid on another traffic demand as well.
\figurename~\ref{fig:rand} shows the same plots with a random traffic demand,
i.e. each distinct pair of hosts sends traffic to each other with a random rate
between 0 and 1. We still see a flat curve for ECMP in fat tree topologies, and
with $k=4$ shortest paths gives a performance close to that of ECMP.

%

\subsection{Non-shortest paths}\label{sub:nonshort}

If we relax our constraint so that all loop free paths can be chosen instead of
just the shortest paths, we found an improvement on irregular topologies.
Consider the curves labelled with $\theta=\infty$ in
\figurename~\ref{fig:const} and \figurename~\ref{fig:rand}, we observe a
lowered maximum link load in those two irregular topologies. This is because,
in irregular topologies, there may not be a large number of parallel shortest
paths. Hence we cannot achieve a well-balanced link load even if we forward
traffic using ECMP, or $k$ shortest paths are used. However, if we relax the
constraint of shortest paths, we may use a longer path to bypass a hot link,
hence the network is better balanced. This explains curves of $\theta=\infty$
in \figurename~\ref{sub:att}--b and \ref{fig:rand}a--b having their right end
lowered. Anyhow, these further shows that ECMP is not always optimal in load
balancing traffic.

Nevertheless, allowing any loop free paths is not without disadvantages.
Besides using a very long path is sometimes not welcomed, consider the fat tree
topologies in \figurename~\ref{fig:const} and \ref{fig:rand}, it is obvious
that using non-shortest path make the network more imbalanced. This is because
in the heuristic algorithm, a path is selected according to a cost function.
Therefore, in the early phase of the algorithm, it is very likely that a long
path is chosen due to a slightly lower load, resulting in a larger sum of loads
across all the links. This is less significant in irregular network because of
the limited path diversity. Thus a longer path is more likely to content with
other traffic, which makes it less likely to be chosen. We found that in
irregular networks, the longer paths are chosen in the later phase of the
algorithm, in contrast to what is done in regular topologies.

We found that using $\theta=25\%$ can combine the advantages. This allows a
slightly longer path, in case the shortest path is too congested, but do not
allow a overly long one. On one hand, this allows a detour around congested
links in irregular network, and on the other hand, enforced shortest path to be
used in fat trees, since the alternative paths is usually doubly long. In
\figurename~\ref{fig:const}-\ref{fig:rand}, we see the curve of $\theta=25\%$
overlaps with that of $\theta=0$ in fat tree topologies and overlaps with that
of $\theta=\infty$ in irregular topologies at the right ends. We justify this
value of $\theta$ by the fact that in data center networks, the usual hop count
in a paths is a handful. Thus allowing 25\% longer essentially means to have
one or two more hops beyond the shortest path, which have limited impact on the
end to end latencies.

\subsection{Robustness to traffic matrix}\label{sub:varied}

Our heuristic algorithm takes a traffic matrix as input. A major criticism on
this approach is the accuracy of the traffic matrix. Not only it is difficult
to measure the traffic matrix, but also it is varying quickly with time. The
algorithm relies on the traffic matrix to select a path. If the paths provided
by the algorithm is very sensitive to the accuracy of the traffic matrix, it
may not be very useful.

\begin{figure}
\subfloat[]{\includegraphics[width=0.5\columnwidth]{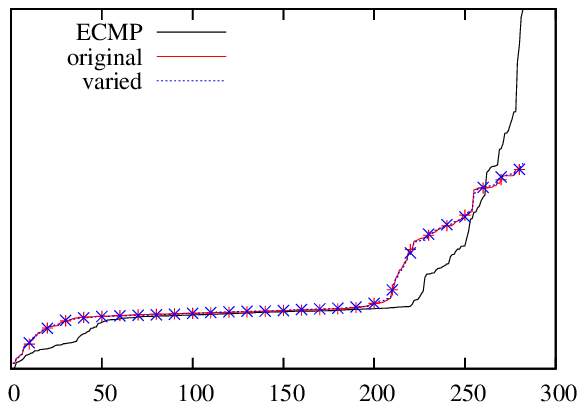}}
\subfloat[]{\includegraphics[width=0.5\columnwidth]{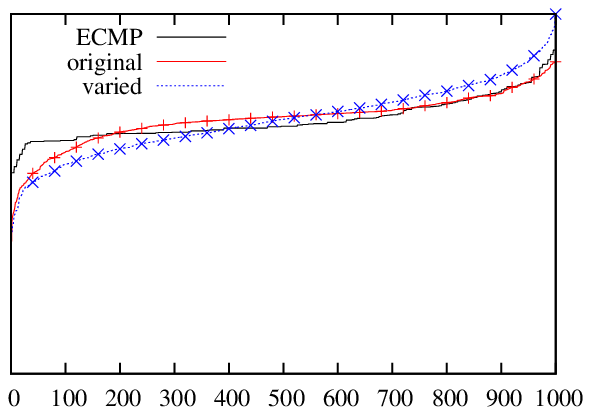}}
\caption{Varied traffic matrix applied to (a) irregular topology 1 and (2) fat tree XGFT(2;5,10;5,5)}\label{fig:var}
\end{figure}

We evaluated the algorithm's sensitivity to traffic matrix by applying a varied
traffic matrix to the paths found by the algorithm. We varied the traffic
matrix by multiplying each entry by a random value between 0.5 and 1.5, then
feed in this traffic matrix to the paths found by the heuristic algorithm with
$k=4$ and $\theta=25\%$. The variation is large, but according to
\figurename~\ref{fig:var}, the link load does not change in scale. Especially
in the irregular network (\figurename~\ref{fig:var}a), the maximum link load is
increased by about 3\%, since the hot link in the irregular network is used by
a lot of source-destination pairs, which the variation between different pairs
are canceling out each other. The fat tree network in
\figurename~\ref{fig:var}b shows around 15\% variation in the maximum link load.
This is because the high degree of path diversity make the variation in
traffic intensity less easy to cancel out. However, the increase in the maximum
link load is still less than the variation in the traffic intensity.  For
reference purposes, we include in \figurename~\ref{fig:var} the link load
according to ECMP when the varied traffic matrix is applied.

\subsection{Effect on the number of paths used}\label{sub:k}

The number of paths $k$ is a parameter to the algorithm. A natural question is
the optimal value of $k$ to be used. In the algorithm, $k$ specifies the
maximum number of paths to be used, which in certain cases, we may not be able
to find $k$ distinct paths for a particular source-destination pair, especially
when the topology does not provide enough path diversity for a particular pair
of nodes. Therefore, we should understand $k$ as the parameter on the
\emph{maximum} number of paths between two nodes.

\begin{figure}
\includegraphics[width=0.5\columnwidth]{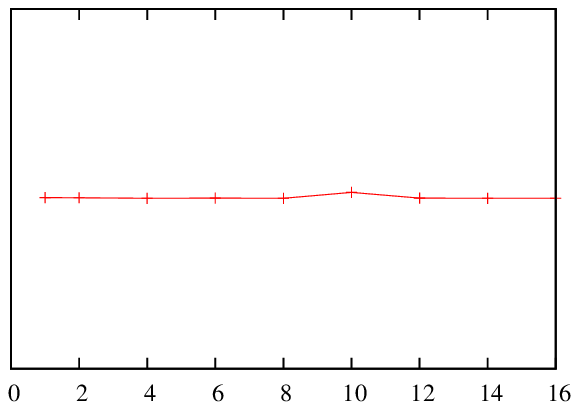}%
\includegraphics[width=0.5\columnwidth]{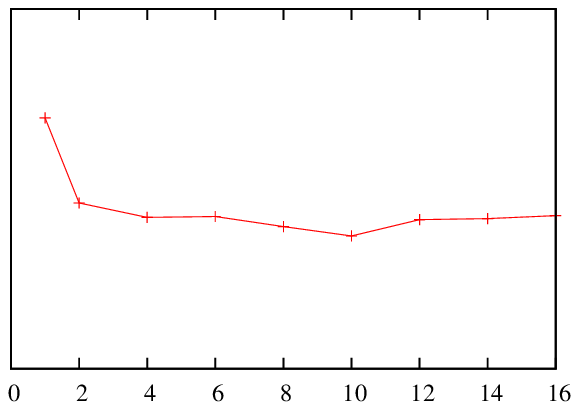}
\caption{Maximum link load vs $k$ under random traffic in (a) irregular topology and (b) XGFT(2;5,10;5,5)}\label{fig:krand}
\end{figure}

\begin{figure}
\includegraphics[width=0.5\columnwidth]{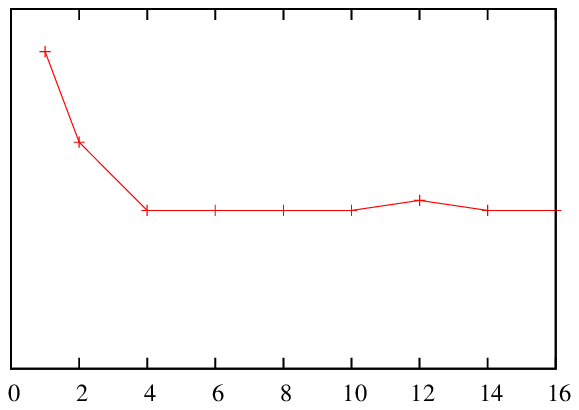}%
\includegraphics[width=0.5\columnwidth]{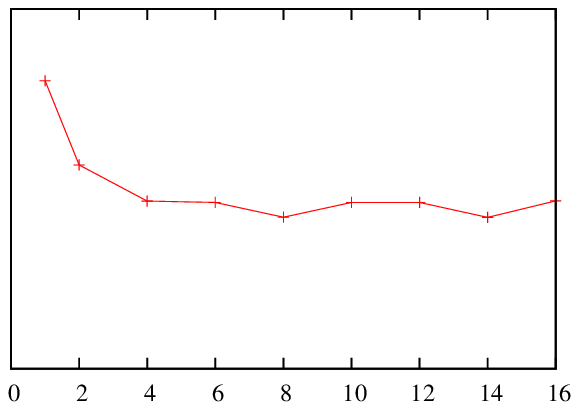}
\caption{Maximum link load vs $k$ under skewed traffic in (a) irregular topology and (b) XGFT(2;5,10;5,5)}\label{fig:kpareto}
\end{figure}

\figurename~\ref{fig:krand} shows the maximum link load in a topology against
$k$ under random traffic. The traffic matrix is the same as the one used in
\figurename~\ref{fig:rand}, and we fix $\theta=25\%$ in the heuristic
algorithm. In \figurename~\ref{fig:krand}b, it is obvious that we can lower the
maximum link load with a larger $k$.  While it is trivial that a larger
$k$ can provide more paths to spread out the traffic and hence lower the
maximum load, it is interesting to see that the curve levels off at a small
value of $k=2$. This can be explained by the fact that the $k$ paths provided
for a pair of nodes may not be mutually disjoint.  Thus further increasing $k$
cannot reduce the load on those shared links by multipath routing.

In \figurename~\ref{fig:krand}a, however, $k$ does not show any effect on the
maximum link load. It is partially due to the topology does not provide enough
path diversity, but also because of the traffic between nodes are even. We
repeated the evaluation with a skewed traffic matrix, i.e. among the nodes in
the topology, we randomly select some of them as hot sender and some as hot
receivers. The cardinality of the set of hot senders and hot receivers are both
20\% of total number of nodes. The traffic from hot senders to hot receivers
comprises 80\% of total traffic across the network. The result is depicted in
\figurename~\ref{fig:kpareto}. Here we see a magnified effect of $k$ to the
maximum link load, but it levels off eventually at a larger value of $k$. An
even more skewed traffic matrix can be constructed. An extreme would be to have
only one flow between two nodes in the network. Then we can expect it can enjoy
a larger $k$ until the paths found have some overlap. But such traffic model
may not be realistic in practice.

From these figures, we see that a small number of $k$ (e.g. $k=4$) is enough to
load balance traffic. Taking the fat tree topology XGFT(2;5,10,5,5) as an
example, there are 25 distinct shortest paths between any two nodes. But
\figurename~\ref{fig:krand}-\ref{fig:kpareto} shows that we need no more than 4
paths from the heuristic algorithm. This is a encouraging result as it means
that a small number of paths is sufficient in practice to achieve a good load
balancing, as a result, the management overhead is limited.

\subsection{Dynamic traffic distribution}\label{sub:dyn}

The above traffic distribution is assuming a fluid traffic that we can
perfectly split across the multiple paths provided. However, usually we do not
want to split a \emph{flow} across different paths to avoid the problems of
packet reordering. Usually a \emph{flow} here is defined as a \emph{five-tuple
flow}, namely, the one identified by the set of source and destination
addresses, protocol, and source and destination port number.  Therefore, we
devised a discrete event simulation to verify the results above is valid in
this setting.

This evaluation is performed as follows: Assume we have the topology and
traffic matrix as before, we can then find the forwarding paths for different
source-destination pairs according to the algorithm. Then, we read the flow
descriptions and place them into the network on one of the those paths found by
the algorithm. The description of one flow provides the information about the
flow's source and destination node, the size (in term of traffic rate of data),
the time of arrival, and its holding time.  We assume it is a CBR flows of
predetermined holding time for simplicity. A flow is placed into the network
one at a time. Each flow uses one path, even if the network provided enough
path diversity, as a means to avoid packet reordering. The path selection for
a particular flow is random. Adaptive placement of flows according to the
real-time load can be a future extension, whereas the random placement can be
regarded as the baseline performance. When a flow is placed to a path, all the
links correspond to this path would have their load increased. Their load will
be decreased when this flow expires. We keep track on the load of each link at
any time.

\begin{figure}
\includegraphics[width=0.5\columnwidth]{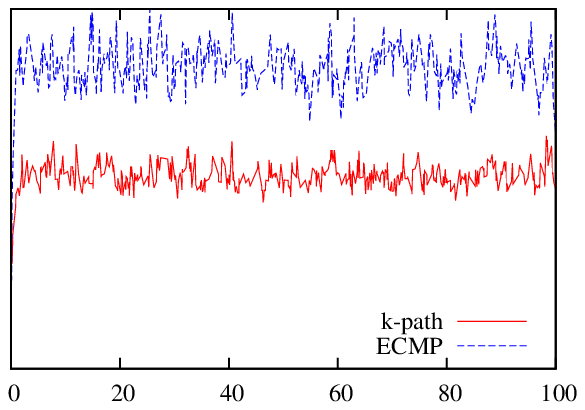}%
\includegraphics[width=0.5\columnwidth]{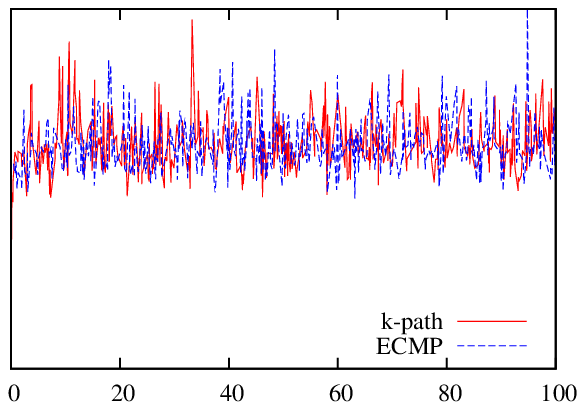}
\caption{Maximum link load vs time under random flows in (a) irregular topology and (b) XGFT(2;5,10;5,5)}\label{fig:dyn}
\end{figure}

\figurename~\ref{fig:dyn} shows the result from the evaluation using dynamic
traffic. The traffic matrix is same as the one used in
\figurename~\ref{fig:rand}. We convert the traffic matrix into a series of
flows with Poisson arrival and exponential holding time, so that their average
load corresponds to the entry in the traffic matrix. The arrival of flows lasts
from time 0 to 100. The maximum link load in the topology against time is
plotted as a continuous curve.

The curve labelled as k-path correspond to the multipath routing according to
the paths given by the algorithm with $k=4$ and $\theta=25\%$.
\figurename~\ref{fig:dyn}b shows that the curves of k-path and ECMP are
overlapping in fat tree topology. This means the former is as good as the
latter. Similarly, \figurename~\ref{fig:dyn}a shows the two curves in an
irregular topology. We can see an observable improvement of the k-path
algorithm over ECMP, in the sense that the former has a lower maximum link load
than the latter. This conclusion is consistent with that of
\figurename~\ref{fig:rand}. For tidiness, we do not show the curves with
$\theta=0$ and $\theta=\infty$ in \figurename~\ref{fig:dyn}, but we confirmed
that their flow-based evaluation is consistent with the abovementioned.

In summary, although we do not consider a flow can be split in this
evaluation, we see the conclusion the same here as in sections
\ref{sub:subset} and \ref{sub:nonshort}, namely, the paths from the algorithm with
$\theta=0$ is as good as ECMP is all scenarios tested, and $\theta=\infty$ can
perform better in irregular topologies where there are a few hot-spot links, or
perform worse in highly regular topologies such as fat-tree. Therefore,
$\theta=25\%$ is suggested to enjoy the benefit in both topologies.

\section{Conclusion}

In this paper, we devised a heuristic algorithm to find a subset of loop free
paths between two nodes in a topology, so that we can approximately minimize
the maximum link load. The heuristic algorithm takes a traffic matrix as the
input to provide estimated measure on the relative traffic intensity between
different pairs of nodes. This algorithm tries to use only a few paths to
forward traffics between different nodes so that contention is minimized. The
output of the algorithm, namely, the set of paths between every pair of nodes,
can be implemented in the network as end-to-end tunnels such as MPLS LSP, to
forward traffic at a later time. The benefit of doing so is to allow the
controllability on the routing of traffic and limit the operation overhead by
limiting the number of paths involved.

We evaluated the algorithm by different traffic models. In the evaluations, we
found that forwarding traffic to all the possible paths is not optimal as the
paths may partially overlap, and thus the link load is imbalanced. Hence this
justifies our approach on using a subset of paths instead of ECMP in term of
load balancing. Furthermore, we found that allowing the traffic to route over
slightly longer paths can improve load balancing in some topologies. If there
are hot links in such topologies, enforcing shortest path routing rules out the
possibility of using a detour to bypass those hot links. Using non-shortest
path is not easy to guarantee the paths are loop free when routing is done
hop-by-hop, but it is trivial when end-to-end tunnels are used. We also found
that the number of paths per pair of nodes is small to achieve a good level of
load balancing. These supports our proposal of using a few tunnels between
pairs of nodes for better control and better load balancing.

\balance
\bibliographystyle{IEEEtran}
\bibliography{ecmp}

\begin{thebibliography}{10}
\providecommand{\url}[1]{#1}
\csname url@samestyle\endcsname
\providecommand{\newblock}{\relax}
\providecommand{\bibinfo}[2]{#2}
\providecommand{\BIBentrySTDinterwordspacing}{\spaceskip=0pt\relax}
\providecommand{\BIBentryALTinterwordstretchfactor}{4}
\providecommand{\BIBentryALTinterwordspacing}{\spaceskip=\fontdimen2\font plus
\BIBentryALTinterwordstretchfactor\fontdimen3\font minus
  \fontdimen4\font\relax}
\providecommand{\BIBforeignlanguage}[2]{{%
\expandafter\ifx\csname l@#1\endcsname\relax
\typeout{** WARNING: IEEEtran.bst: No hyphenation pattern has been}%
\typeout{** loaded for the language `#1'. Using the pattern for}%
\typeout{** the default language instead.}%
\else
\language=\csname l@#1\endcsname
\fi
#2}}
\providecommand{\BIBdecl}{\relax}
\BIBdecl

\bibitem{rfc2328}
J.~Moy, ``{OSPF} version 2,'' IETF STD 54, RFC 2328, Apr. 1998.

\bibitem{rfc3031}
E.~C. Rosen, A.~Viswanathan, and R.~Callon, ``Multiprotocol label switching
  architecture,'' IETF RFC 3031, Jan. 2001.

\bibitem{myam10}
J.~Mudigonda, P.~Yalagandula, M.~Al-Fares, and J.~C. Mogul, ``{SPAIN}: {COTS}
  data-center {Ethernet} for multipathing over arbitrary topologies,'' in
  \emph{Proc. 7th USENIX NSDI}, San Jose, CA, Apr. 2010.

\bibitem{ejlw01}
A.~Elwalid, C.~Jin, S.~Low, and I.~Widjaja, ``{MATE}: {MPLS} adaptive traffic
  engineering,'' in \emph{Proc. INFOCOM}, 2001.

\bibitem{mabpprst08}
N.~McKeown, T.~Anderson, H.~Balakrishnan, G.~Prulkar, L.~Peterson, J.~Rexford,
  S.~Shenker, and J.~Turner, ``{OpenFlow}: Enabling innovation in campus
  networks,'' \emph{ACM SIGCOMM Computer Communication Review}, vol.~38, no.~2,
  pp. 69--74, Apr. 2008.

\bibitem{baaz10}
T.~Benson, A.~Anand, A.~Akella, and M.~Zhang, ``The case for fine-grained
  traffic engineering in data-centers,'' in \emph{Proc. INM/WREN}, San Jose,
  CA, Apr. 2010.

\bibitem{ns99}
P.~Narvaez and K.~Y. Siu, ``Efficient algorithms for multi-path link state
  routing,'' in \emph{Proc. ISCOM}, 1999.

\bibitem{vg00}
S.~Vutukurya and J.~Garcia-Luna-Aceves, ``{MPATH}: a loop-free multipath
  routing algorithm,'' \emph{Microprocessors and Microsystems}, vol.~24, pp.
  319--327, 2000.

\bibitem{vg01}
S.~Vutukury and J.~J. Garcia-Luna-Aceves, ``{MDVA}: A distance-vector multipath
  routing protocol,'' in \emph{Proc. INFOCOM}, 2001.

\bibitem{mpc08}
P.~M\'{e}rindol, J.-J. Pansiot, and S.~Cateloin, ``Improving load balancing
  with multipath routing,'' in \emph{Proc. ICCCN}, 2008, pp. 1--8.

\bibitem{ft00}
B.~Fortz and M.~Thorup, ``Internet traffic engineering by optimizing {OSPF}
  weights,'' in \emph{Proc. INFOCOM}, 2000, pp. 519--528.

\bibitem{bo07}
R.~Banner and A.~Orda, ``Multipath routing algorithms for congestion
  minimization,'' \emph{IEEE/ACM Transactions on Networking}, vol.~15, no.~2,
  pp. 413--424, Apr. 2007.

\bibitem{wc96}
Z.~Wang and J.~Crowcroft, ``Quality-of-service routing for supporting
  multimedia applications,'' \emph{IEEE Journal on Selected Areas in
  Communications}, vol.~14, no.~7, pp. 1228--1234, 1996.

\bibitem{ms97}
Q.~Ma and P.~Steenkiste, ``On path selection for traffic with bandwidth
  guarantees,'' in \emph{Proc. ICNP}, Oct. 1997.

\bibitem{rfc2386}
R.~Guerin, S.~Kamat, A.~Orda, T.~Przygienda, and D.~Wiliams, ``Qos routing
  mechanisms and ospf extensions,'' IETF RFC 2386, Aug. 1998.

\bibitem{cn98}
S.~Chen and K.~Nahrstedt, ``An overview of quality-of-service routing for the
  next generation high-speed networks: Problems and solutions,'' \emph{IEEE
  Network Magazine}, vol.~12, no.~6, pp. 64--79, 1998.

\bibitem{jng01}
Y.~Jia, I.~Nikolaidis, and P.~Gburzynski, ``Multiple path {QoS} routing,'' in
  \emph{Proc. ICC}, 2001, pp. 2583--2587.

\bibitem{ospfomp}
C.~Villamizar, ``{OSPF} optimized multipath ({OSPF-OMP}),'' IETF Internet
  Draft, Feb. 1999, \url{http://tools.ietf.org/html/draft-ietf-ospf-omp-02}.

\bibitem{mplsomp}
------, ``{MPLS} optimized multipath ({MPLS-OMP}),'' IETF Internet Draft, Feb.
  1999, \url{http://tools.ietf.org/html/draft-villamizar-mpls-omp-01}.

\bibitem{nzd04}
S.~Nelakuditi, Z.-L. Zhang, and D.~H. Du, ``On selection of candidate paths for
  proportional routing,'' \emph{Computer Networks}, vol.~44, pp. 79--102, 2004.

\bibitem{gjkklmps09}
A.~Greenberg, N.~Jain, S.~Kandula, C.~Kim, P.~Lahiri, D.~A. Maltz, P.~Patel,
  and S.~Sengupta, ``{VL2}: A scalable and flexible data center network,'' in
  \emph{Proceedings of the ACM SIGCOMM Conference on Data Communication}, Aug.
  2009.

\bibitem{gzrr03}
I.~Gojmerac, T.~Ziegler, F.~Ricciato, and P.~Reichl, ``Adaptive multipath
  routing for dynamic traffic engineering,'' in \emph{Proc. GLOBECOM}, Nov.
  2003.

\bibitem{pm04}
M.~Pi\'{o}ro and D.~Medhi, \emph{Routing, Flow, and Capacity Design in
  Communication and Computer Networks}.\hskip 1em plus 0.5em minus 0.4em\relax
  San Francisco, CA: Morgan Kaufmann, 2004.

\bibitem{y71}
J.~Y. Yen, ``Finding the $k$ shortest loopless paths in a network,''
  \emph{Management Science}, vol.~17, no.~11, pp. 712--716, Jul. 1971.

\bibitem{m78}
E.~Minieka, \emph{Optimization algorithms for networks and graphs}.\hskip 1em
  plus 0.5em minus 0.4em\relax New York: Marcel Dekker, 1978.

\bibitem{e98}
D.~Eppstein, ``Finding the $k$ shortest paths,'' \emph{SIAM J. Computing},
  vol.~28, no.~2, pp. 652--673, 1998.

\end{thebibliography}

\label{docend}
\end{document}